\documentclass[reprint,
superscriptaddress,
 amsmath,amssymb,
 aps,
]{revtex4-1}

\usepackage{color}
\usepackage{graphicx}
\usepackage{dcolumn}
\usepackage{bm}
\usepackage{graphicx}
\usepackage{epstopdf}
\usepackage{gensymb}
\usepackage{braket}

\usepackage{float}

\begin{document}

\title{Structure and magnetism in Fe-doped FeVSb and epitaxial Fe/FeVSb nanocomposite films}

\author{Estiaque H. Shourov}
\affiliation{Materials Science and Engineering, University of Wisconsin--Madison}

\author{Chenyu Zhang}
\affiliation{Materials Science and Engineering, University of Wisconsin--Madison}

\author{Paul M. Voyles}
\affiliation{Materials Science and Engineering, University of Wisconsin--Madison}

\author{Jason K. Kawasaki}
\email{jkawasaki@wisc.edu}
\affiliation{Materials Science and Engineering, University of Wisconsin--Madison}

\date{\today}
\begin{abstract}

The combination of ferromagnetism and semiconducting behavior offers an avenue for realizing novel spintronics and spin-enhanced thermoelectrics. Here we demonstrate the synthesis of doped and nanocomposite half Heusler Fe$_{1+x}$VSb films by molecular beam epitaxy. For dilute excess Fe ($x < 0.1$), we observe a decrease in the Hall electron concentration and no secondary phases in X-ray diffraction, consistent with Fe doping into FeVSb. Magnetotransport measurements suggest weak ferromagnetism that onsets at a temperature of $T_{c} \approx$ 5K. For higher Fe content ($x > 0.1$), ferromagnetic Fe nanostructures precipitate from the semiconducting FeVSb matrix. The Fe/FeVSb interfaces are epitaxial, as observed by transmission electron microscopy  and X-ray diffraction. Magnetotransport measurements suggest proximity-induced magnetism in the FeVSb, from the Fe/FeVSb interfaces, at an onset temperature of $T_{c} \approx$ 20K.

\end{abstract}

\maketitle

\section{Introduction}

Incorporating magnetism and epitaxial interfaces in semiconducting half Heusler compounds is attractive for applications in spintronics and thermoelectric power conversion. While half-Heusler compounds with 18 valence electrons per formula unit are generally diamagnetic semiconductors \cite{jung2000study,kandpal2006covalent}, slight deviations from stoichiometry can make these materials magnetic \cite{tobola1998crossover, harrington2018growth}. This provides a route to make new dilute magnetic semiconductors for applications in spintronics \cite{casper2012half, wollmann2017heusler}. Half-Heusler compounds are also attractive thermoelectric materials due to their large thermoelectric power factors \cite{young2000thermoelectric} and the ability to precipitate nanostructures to decrease the thermal conductivity \cite{douglas2012enhanced}. New concepts based on magnon drag \cite{vandaele2017thermal} and spin fluctuations \cite{tsujii2019observation} suggest that incorporating magnetism may further increase the thermopower.
 
Here we explore the structure and magnetism of epitaxial thin films with total composition Fe$_{1+x}$VSb, grown by molecular beam epitaxy (MBE) on MgO (001) substrates. FeVSb is a semiconducting, 18 valence electron half Heusler compound. For dilute $x$, excess Fe is expected to incorporate into the FeVSb lattice and make this compound ferromagnetic \cite{tobola2000electronic}. Additionally, FeVSb and Fe (body centered cubic) share a tie line in the Fe-V-Sb ternary phase diagram \cite{romaka2012interaction} and similar lattice parameters ($a_{FeVSb} = 5.82$ \AA, $2a_{Fe,bcc}=5.73$ \AA), enabling the formation of thermodynamically stable, epitaxial Fe:FeVSb nanocomposites for larger values of $x$. 

We show that for Fe$_{1+x}$VSb epitaxial films with $x < 0.1$, the Fe dopes into FeVSb and has a ferromagnetic onset temperature of $T_{c} \approx$ 5K. For $x > 0.1$ we observe epitaxial Fe precipitates embedded within a FeVSb matrix. In these Fe:FeVSb nanocomposites we identify two sources of magnetism: the ferromagnetic Fe precipitates ($T_{c} \gg$ 300K) and proximity-induced ferromagnetism in the FeVSb, from the Fe/FeVSb interfaces. Our work identifies a clean system for exploring magnetic doping, epitaxial nanostructuring, and magnetic proximity effects in thermoelectric and spintronic materials.

\section{Results and Discussion}

\begin{figure*}
    \centering
    \includegraphics[width=0.98\textwidth]{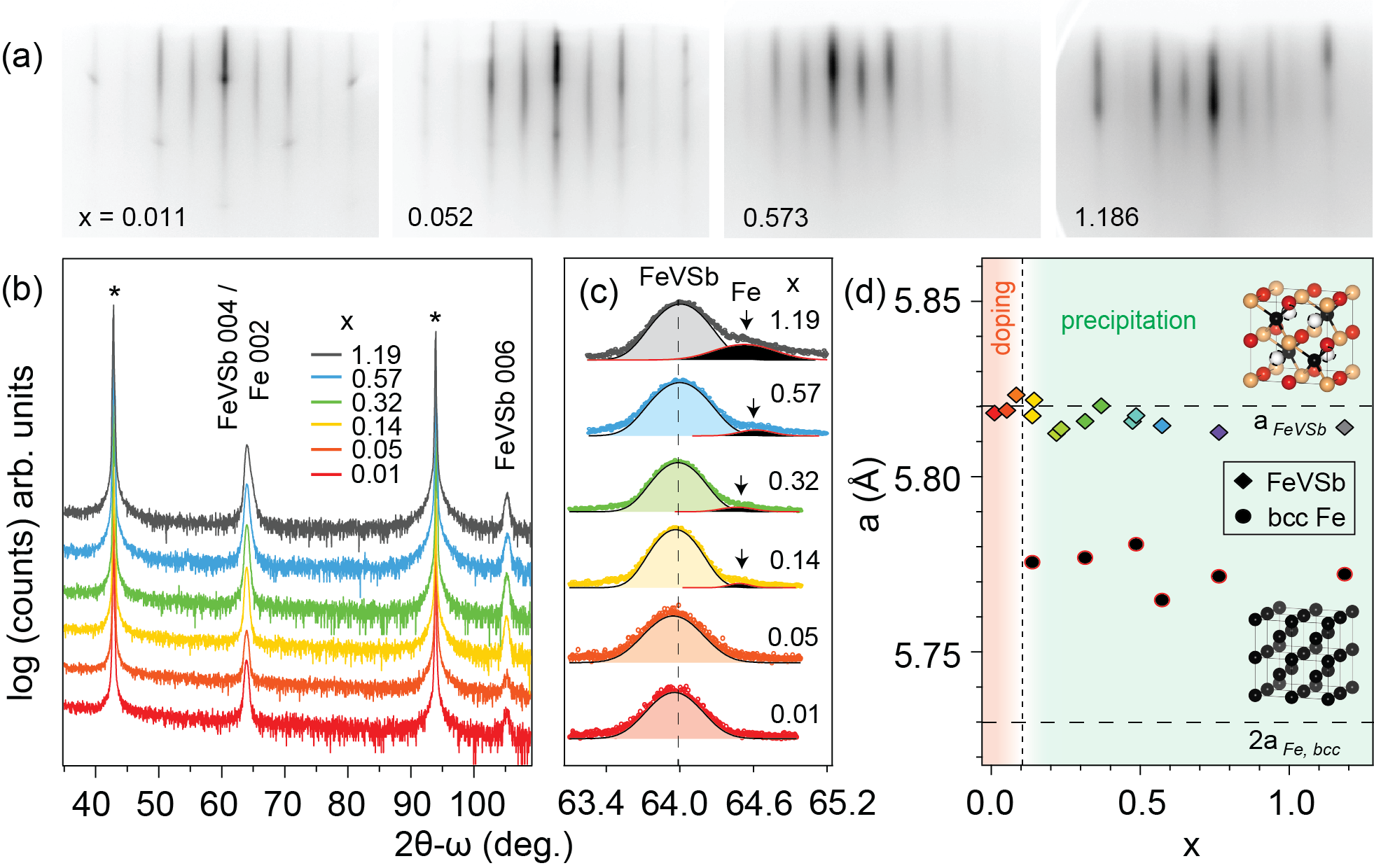}
    \caption{\textbf{Structural evolution of epitaxial Fe$_{1+x}$VSb films by electron and x-ray diffraction.} (a) RHEED pattern along the $<110>$ azimuth, showing strong streaky  2$\times$ reconstruction over all compositions studied. (b) Wide angle XRD (Cu $K\alpha$) showing the half-Heusler 00$l$ and Fe 002 reflections. Asterisks indicate the MgO substrate reflections. (c) High resolution scans of the FeVSb 004 reflections reveal the onset of a shoulder peak at composition $x = 0.14$, which we attribute to the 002 reflection of Fe (bcc). Shaded curves show the Gaussian fits. (d) Out of plane lattice parameter extracted from XRD as a function of excess Fe composition. Diamond and circle markers correspond to FeVSb and Fe respectively. Dotted lines show the lattice parameter of bulk FeVSb (half Heusler), and that of doubled body centered cubic Fe unit cells. Crystal structure models for FeVSb and Fe are shown. Black, red, orange and white spheres corresponds to Fe, V, Sb and interstitial respectively. For low $x$, excess Fe is expected to incorporate into the ($\frac{3}{4}, \frac{1}{4}, \frac{1}{4}$) tetrahedral interstitial sites of FeVSb (white spheres).}
    \label{structure}
\end{figure*}

Fe$_{1+x}$VSb films with varying $x$ were grown by molecular beam epitaxy (MBE) on MgO (001) substrates. Samples were grown leveraging a semi-adsorption controlled growth window in which the Sb stoichiometry is self-limiting \cite{shourov2020semi}. The Fe and V fluxes were measured \textit{in situ} using a quartz crystal microbalance (QCM) immediately prior to sample growth. Absolute compositions were calibrated using Rutherford Backscattering Spectrometry (RBS) on separate samples. Further details on the growth process can be found elsewhere \cite{shourov2020semi}. 

\begin{figure}
    \centering
    \includegraphics[width=0.45\textwidth]{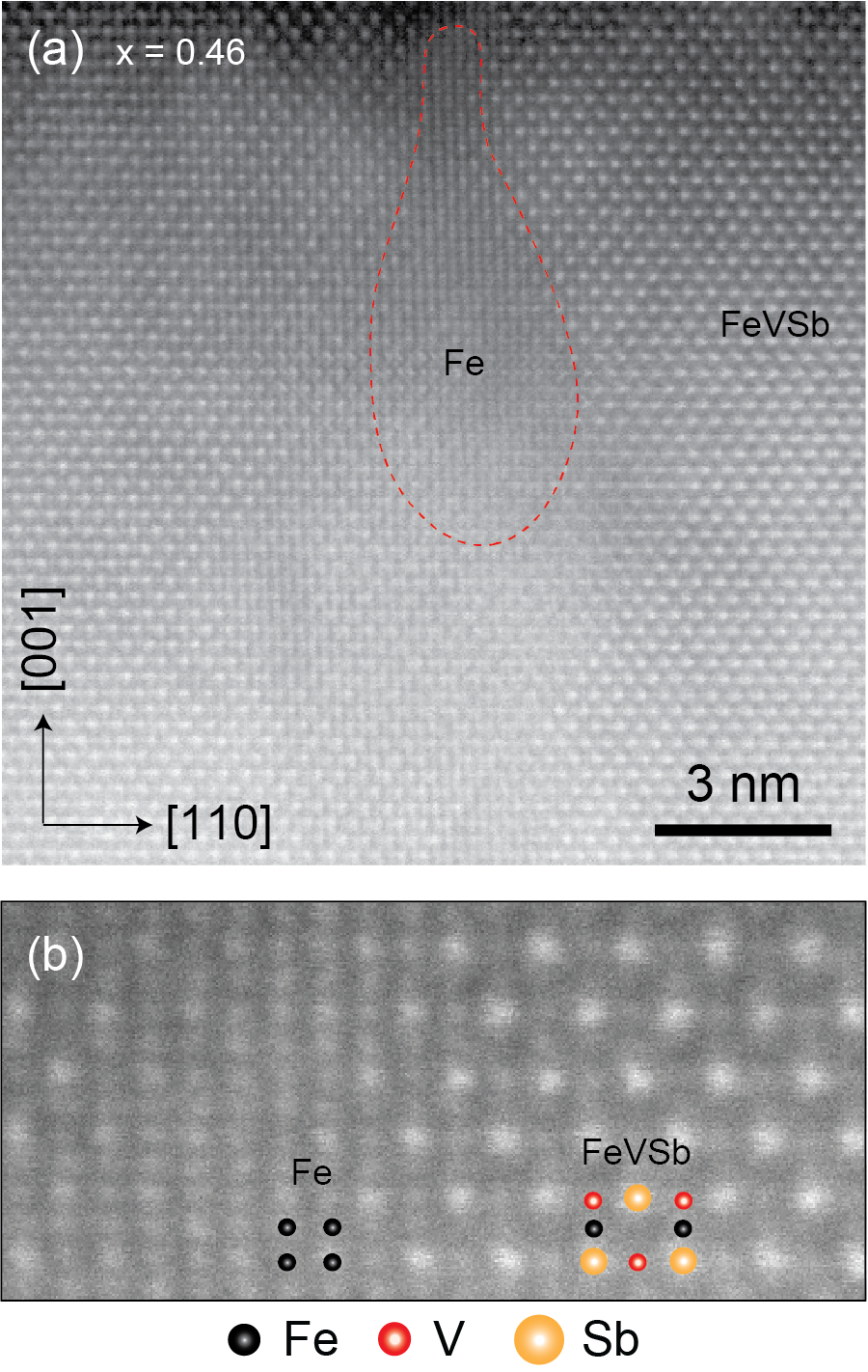}
    \caption{\textbf{Cross-sectional STEM image showing Fe precipitation in the Fe$_{1.46}$VSb film.} (a) HAADF-STEM image of nanometer scale Fe nanoprecipitates embedded in the FeVSb matrix. (b) High resolution image of Fe/FeVSb interface. Individual atoms are identified by atomic models, confirming the half-Heusler and \textit{bcc} crystal structures of the two phases.}
    \label{tem}
\end{figure}

In Fig. \ref{structure}, we investigate the structural evolution of the films as a function of excess Fe. In the reflection high energy electron diffraction (RHEED, Fig. \ref{structure}(a)), all films display a characteristic 2$\times$ streaky reconstructed surface indicative of smooth epitaxial films. The 2$\times$ reconstruction is attributed to Sb dimerization which is well-known for antimonide half-Heusler surfaces \cite{kawasaki2018simple}. 

X-ray diffraction (XRD) confirms that the films are all epitaxial. In the wide angle $2\theta-\omega$ scan (Fig. \ref{structure}(b)), only $00l$-type FeVSb (half-Heusler) and Fe (bcc) reflections are observed. For $x < 0.1$, high resolution scans around the FeVSb 004 reflection detect only the FeVSb half Heusler phase (Fig. \ref{structure}c). In this dilute regime, we expect the excess Fe to occupy the tetrahedral interstitial ($\frac{3}{4}, \frac{1}{4}, \frac{1}{4}$) sites in the FeVSb lattice \cite{yonggang2017natural} (Fig. \ref{structure}d, white spheres). For $x < 0.1$, while we do not observe a Fe bcc 002 reflection, it is possible that some bcc Fe phase is present, below the detection limit of XRD. For $x \geq 0.14$, we observe a secondary peak at $2\theta = 64.5$ degrees, which we attribute to the 002 reflection of body centered cubic Fe. The secondary peak grows in intensity with increasing $x$, which we attribute to an increasing volume fraction of Fe precipitates. The lattice parameters ($a$) calculated from each peak are plotted in Fig. \ref{structure}(d). The primary peak lattice parameter agrees well to that of bulk FeVSb. The secondary peak lattice parameter appears to match a dilated doubled unit cell of bcc Fe. We attribute the slight increase in lattice parameter of Fe to strain. 

Scanning transmission electron microscopy (STEM) for samples with $x \geq 0.1$ confirms the existence of bcc Fe precipitates, embedded epitaxially within a FeVSb matrix. Fig. \ref{tem} shows a high angle annular dark field (HAADF) STEM image of a sample with $x = 0.46$, in which we identify a Fe precipitate. Closer analysis of the Fe/FeVSb interface reveals a cube on cube epitaxial relationship, with Fe(001)[110] $\parallel$ FeVSb(001)[110]. We have identified Fe precipitates in Fe$_{1+x}$VSb samples with $x$ as small as 0.1 (Supplemental Fig. S-1). No Fe precipitates have been identified for samples with $x < 0.1$. Further structural analysis is required to more precisely determine the Fe solubility limit in FeVSb.

\begin{figure}[h]
    \centering
    \includegraphics[width=0.4\textwidth]{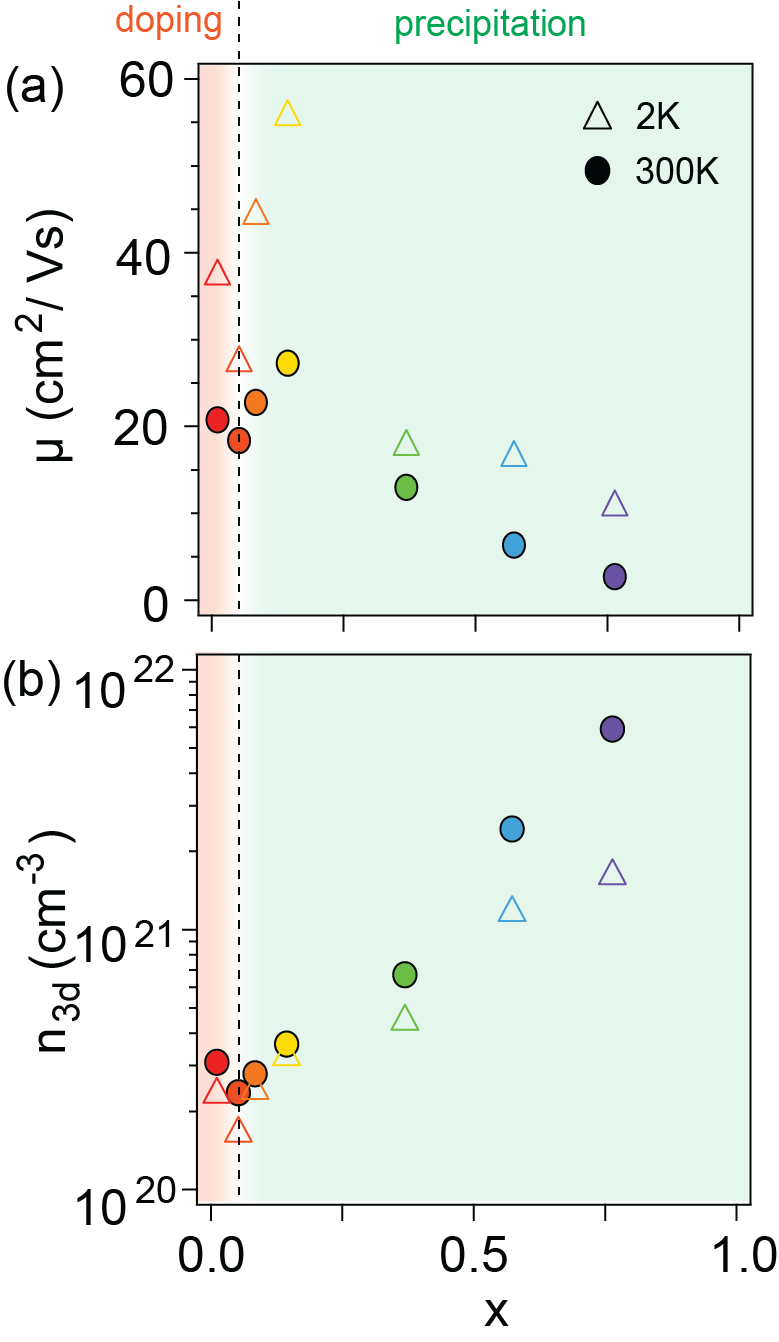}
    \caption{\textbf{Transport measurements for films with varying Fe content.} (a) Mobility $\mu$ and (b) electron concentration $n_{3d}$ as a function of excess Fe composition $x$ at 300K and at 2K.}
    \label{transport}
\end{figure}

The dependence of the Hall mobility ($\mu$) and carrier concentration ($n_{3d}$) on composition $x$ provides an additional estimate of the solubility limit for excess Fe. Fig. \ref{transport} shows the mobility and carrier concentration extracted from Hall effect measurements. Since the samples with $x > 0$ are ferromagnetic and exhibit contributions from the anomalous Hall effect, we use a linear fit of $\rho_{xy}(B)$ at high field to extract the majority carrier concentration (Supplemental). We find that as a function of $x$, the electron concentration exhibits a minimum near $x \approx 0.05-0.1$ and the mobility exhibits a maximum near $x \approx 0.1-0.14$. These findings suggest that for low $x < 0.1$, the excess Fe acts to compensate free carriers in the FeVSb. Similar results have been observed experimentally for excess Ni in NiTiSn \cite{rice2017structural}. This implies that at low $x < 0.1$, the excess Fe may dope into the FeVSb lattice. First principles calculations suggest the most likely dopant site is the ($\frac{3}{4}, \frac{1}{4}, \frac{1}{4}$)  vacancies in the half-Heusler lattice \cite{yonggang2017natural}.

\begin{figure}[h]
    \centering
    \includegraphics[width=0.48\textwidth]{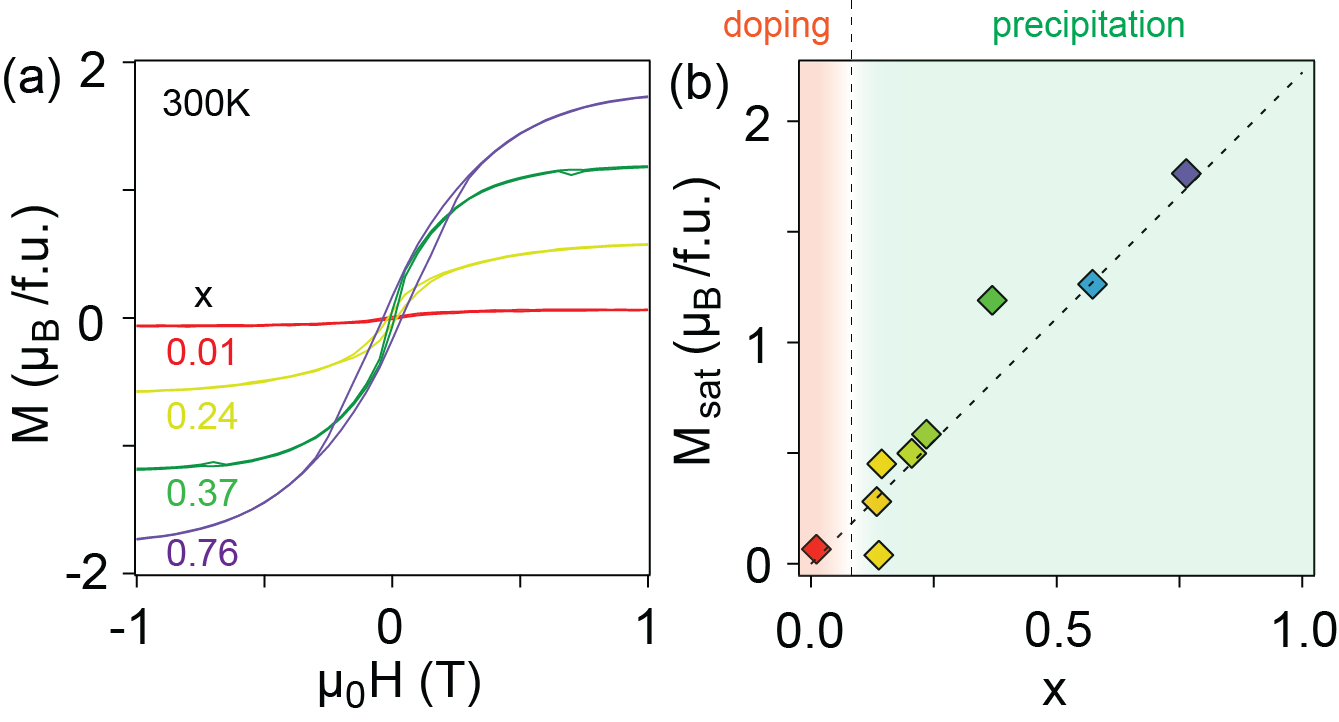}
    \caption{\textbf{Magnetism from precipitated Fe nanoparticles, as measured by SQUID.} (a) Magnetization $M(H)$ for films with varying excess Fe composition. The diamagnetic contribution from the substrate has been subtracted. $H$ was applied out of the sample plane ($H \parallel$ [001]). (b) Saturation magnetization, M$_{sat}$ as a function of $x$. The linear dependence of M$_{sat}$ on $X_{Fe}$ follows the Slater-Pauling curve for bcc Fe, as marked by the dotted line. Magnetization is expressed in units of Bohr magneton ($\mu_{B}$) per formula unit (f.u.) of FeVSb.} 
    \label{squid}
\end{figure}

For films with $x > 0.1$, SQUID magnetometry measurements reveal well defined hysteresis in the magnetization ($M$) versus applied field ($H$) at room temperature, consistent with ferromagnetic Fe precipitates (Fig. \ref{squid}(a)). The magnetization curves are nearly temperature independent in the range from 300K to 50K, suggesting that the Curie temperature of the observed ferromagnetic ordering is much higher than 300K (supplement Fig. 4). The high $T_{c}$ suggests that the ferromagnetism detected by SQUID is dominated by the ferromagnetic Fe precipitates. In contrast, dilute magnetic doping in FeVSb and ferromagnetic proximity effects at Fe/FeVSb interfaces are expected to onset at lower temperatures and have a weaker magnetic response. Fig. \ref{squid}(b) tracks the saturation magnetization ($M_{sat}$) at 300K with the excess Fe content. For $ x > 0.1$ (above the solubility limit), $M_{sat}$ dependence on $x$ follows the linear Slater-Pauling behavior expected for bcc Fe, suggesting that the total magnetization is dominated by the local moment of precipitated Fe nanoparticles. 

\begin{figure*}[t]
    \centering
    \includegraphics[width=1\textwidth]{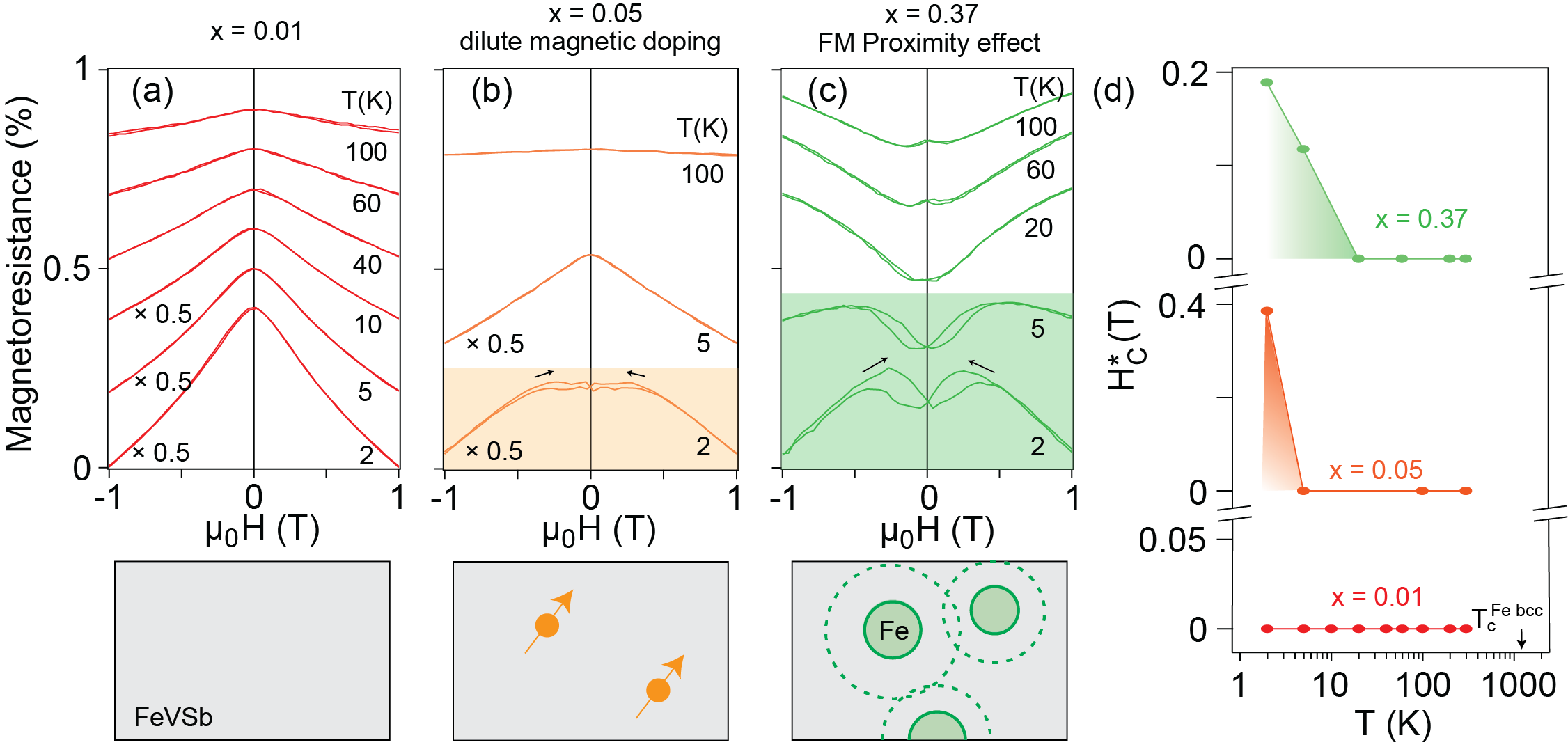}
    \caption{\textbf{Dilute magnetism and ferromagnetic proximity effect in the FeVSb matrix.} (a-c) Magnetoresistance $[\rho_{xx}(H) - \rho_{xx}(0)]/\rho_{xx}(0)$ for Fe$_{1+x}$VSb samples with $x =$ 0.01, 0.05, and 0.37. (d) Width of the magnetoresistance hysteresis as a function of temperature. The $T_{c}$ for bulk Fe is from Ref. \cite{rosengaard1997finite}.}
    \label{new}
\end{figure*}

To understand possible magnetism induced in the FeVSb, especially in the low $x$ limit, we investigate the temperature-dependent magnetoresistance $\Delta \rho_{xx}(H)/\rho_{xx}(0)$. Unlike magnetometry, which detects the sum of all moments in the sample, we expect magnetoresistance to be dominated by the FeVSb matrix, since FeVSb forms a continuous conduction path. The Fe precipitates, in contrast, are disconnected. Fig. \ref{new}(a-c) shows the temperature dependence of the magnetoresistance for samples with varying $x$. For $x = 0.01$, near stoichiometric condition, the magnetoresistance is negative and the magnetoresistance curve exhibits a broad zero field peak. We attribute the broad peak to localization from magnetic polarons \cite{schmidt1999fluctuation} due to slight nonstoichiometry. Similar behavior has been observed for the half-Heusler compound CoTiSb \cite{kawasaki2014growth}. For $x =0.05$ and 0.37, we observe butterfly shaped hysteresis at low temperature indicative of ferromagnetic ordering \cite{ziese2002extrinsic}.

We track the low temperature onset of ferromagnetism by extracting the width of the magnetoresistance. We define the effective coercive field $H_{C}^{*}$ as the field separation between the minima of the $\Delta \rho_{xx}(H)/\rho_{xx}(0)$ curves (supplement Fig. S-5). We define the temperature at which $H_{C}^{*}$ goes to zero as the Curie temperature $T_{c}$. Fig. \ref{new}(d) tracks $H_{C}^{*}(T)$ for the same samples in Fig. \ref{new}(a-c). For $x = 0.05$, which is below the solubility limit, we find that the $T_{c}$ is less than 5 K. We interpret this $T_{c}$ to mark the onset of dilute magnetism in the FeVSb semiconductor. For $x = 0.37$, which is above the solubility limit, we find $T_{c}\sim 20$ K. We interpret this $T_{c}$ to mark the onset of proximity-induced ferromagnetism in the FeVSb, from the Fe/FeVSb interfaces. Both $T_{c}$s are significantly smaller than the $T_{c}$ for Fe precipitates (1043 K  \cite{rosengaard1997finite}).

\section{Conclusion}

In summary, we have demonstrated the MBE synthesis of epitaxial Fe$_{1+x}$VSb films and provided an estimate for the solubility limit of Fe in FeVSb under these (non-equilibrium) growth conditions. We find that for $x < 0.1$ excess Fe appears to incorporate homogeneously into the lattice. Hysteresis in the low temperature magnetoresistance suggests dilute magnetic semiconducting behavior. Above the Fe solubility limit ($x > 0.1$), epitaxial Fe precipitates form, embedded within the FeVSb semiconducting matrix. Two distinct sources of magnetism are identified for Fe:FeVSb nanocomposites: (1) ferromagetism of Fe nanoparticles detected in magnetometry and (2) proximity effect induced magnetism in FeVSb matrix detected in the magnetoresistance measurements. Our work provide a clean platform for the study of magnetic, nanocomposite Heusler systems. 

\section{Acknowledgments}

This work was supported by the CAREER program of the National Science Foundation (DMR-1752797) and by the SEED program of the Wisconsin Materials Research Science and Engineering Center, an NSF funded center (DMR-1720415). We gratefully acknowledge the use of X-ray diffraction and electron microscopy facilities supported by the NSF through the University of Wisconsin Materials Research Science and Engineering Center under Grant No. DMR-1720415. We thank Professor Song Jin for the use of PPMS facilities. The Quantum Design MPMS3 magnetometer was supported by the UW Madison Department of Chemistry. We thank Greg Haugstad (Characterization Facility, University of Minnesota) for performing RBS measurements.

\bibliographystyle{apsrev}
\bibliography{bibliography}

\end{document}